\documentclass[aps,pre,preprint,groupedaddress,showpacs]{revtex4}
\usepackage{graphicx}
\begin{document}
\newcommand{\ud}{{\mathrm d}}
\newcommand{\umod}{\mathrm{mod}}
\newcommand{\sech}{\mathrm{sech}}

\title{Nonlinear Stochastic Resonance with subthreshold  rectangular
pulses}


\author{Jes\'us Casado-Pascual}
\email[]{jcasado@us.es}
\homepage[]{http://numerix.us.es}
\author{Jos\'e  G\'omez-Ord\'o\~nez}
\author{ Manuel Morillo}
\affiliation{F\'{\i}sica Te\'orica,
Universidad de Sevilla, Apartado de Correos 1065, Sevilla 41080, Spain}

\date{\today}

\begin{abstract}
We analyze the phenomenon of nonlinear stochastic resonance (SR) in
noisy bistable systems driven by pulsed time periodic forces. The
driving force contains, within each period, two pulses of equal constant
amplitude and duration but opposite signs. Each pulse starts every
half-period and its duration is varied. For {\em subthreshold}
amplitudes, we study the dependence of the output signal-to-noise ratio
(SNR) and the SR gain on the noise strength and the relative duration of
the pulses.  We find that the SR gains can reach values larger than
unity, with maximum values showing a nonmonotonic dependence on the
duration of the pulses.
\end{abstract}


\maketitle
In recent work \cite{Casgomprl,Casgompre}, we have carried out a
detailed analytical and numerical study of the nonlinear response of a
noisy bistable system subject to a subthreshold time periodic driving
force. The force considered in those works is such that it remains
constant within the duration of each half period while switching its
sign every half period.  We have focused our attention on the analysis
of the dependence of the output signal-to-noise ratio and the
corresponding SR gain on the noise strength.  The analytical study was
based on a two-state approximation amenable to exact treatment
\cite{Casgomprl}.  We showed that, for subthreshold input signals of
sufficiently long periods, the phenomenon of SR can be accompanied by
SR gains larger than unity. This is a genuine characterization of
nonlinearity, as SR gains larger than unity are strictly forbidden
within a linear response description \cite{dykluc95,dewbia95,casgom03}.
The analytical results were corroborated by numerical simulations
\cite{Casgompre}.

In the last few years, Gingl and collaborators
\cite{ginmak00,ginmak01,ginmak02} have carried out analog simulations of
noisy bistable systems subject to driving forces of the type given in
Eq.~(\ref{pulse})
\begin{equation}
\label{pulse}
F(t)= \left \{ \begin{array}
{r@{\quad:\quad}l}
A & 0\le t < t_c \\
-A& \frac T2 \le t < \frac T2 + t_c \\
0& \textrm{otherwise}
\end{array}
\right .
\end{equation}
It is convenient to introduce the parameter $r=2t_c/T$, measuring the
fraction of a period during which this driving force has a nonvanishing
value (the parameter $r$ in the present paper corresponds exactly to
what Gingl and coworkers term ``duty cyle''). In
\cite{ginmak00,ginmak01,ginmak02}, the SNR and the SR gain for
subthreshold amplitude input signals with $r \le 0.3$ were
studied. These authors find SR gains larger than unity, and, also, that
increasing the $r$ value lowers the SNR gain. They rationalize their
observations by noting that the input SNR increases as $r$ increases,
while the output SNR is less sensitive to the value of $r$. The case
studied by us in \cite{Casgomprl,Casgompre} corresponds to the largest
possible value of the parameter $r$, namely, $r=1$.  It seems therefore
of interest to extend our analysis to input signals with $r < 1$ in
order to compare with the predictions of Gingl et al.

Let us consider a system characterized by a single degree of freedom,
$x$, whose dynamics (in dimensionless units)  is governed by the
Langevin equation
\begin{equation}
\label{langev} \dot{x}(t)=-U'\left[ x(t),t \right]+\xi(t),
\end{equation}
where $\xi (t)$ is a Gaussian white noise of zero mean with $\langle
\xi(t)\xi(s)\rangle = 2D\delta(t-s)$, and $ -U'(x,t)$ represents the
force stemming from the time-dependent, archetype bistable potential
\begin{equation}
\label{potential}
U(x,t)=\frac{x^4}{4}-\frac{x^2}{2}-F(t)x,
\end{equation}
with $F(t)$ given by Eq.~(\ref{pulse}). The one-time correlation
function is defined as
\begin{equation}
\label{ctau}
C(\tau)= \frac{1}{T} \int_{0}^{T} \ud t \, \langle x(t+\tau)
x(t)\rangle_{\infty}.
\end{equation}
It can be written exactly as the sum of two contributions: a coherent
part, $C_{coh}(\tau)$, which is periodic in $\tau$ with period $T$, and
an incoherent part, $C_{incoh}(\tau)$, which decays to $0$ for large
values of $\tau$ and reflects the correlation of the output fluctuations
about its average (the noisy part of the output).  The coherent
part $C_{coh}(\tau)$ is given by \cite{RMP,PRA91}
\begin{equation}
\label{chtau}
C_{coh}(\tau) =\frac{1}{T} \int_{0}^{T} \ud t \,\langle x(t+\tau)
\rangle_{\infty} \langle x(t) \rangle_{\infty},
\end{equation}
and $C_{incoh}(\tau)$ is obtained from the difference of
Eq.~(\ref{ctau}) and Eq.~(\ref{chtau}).  In the expressions above, the
susbcript indicates that the averages are to be evaluated in the limit
$t \rightarrow \infty$.  

According to McNamara and Wiesenfeld \cite{McNWie89}, the output SNR,
$R_{out}$, is defined in terms of the Fourier transform of the
coherent and incoherent parts of $C(\tau)$ as
\begin{equation}
\label{snr}
R_{out} =\frac {\lim_{\epsilon \rightarrow 0+}
\int_{\Omega-\epsilon}^{\Omega+\epsilon} \ud\omega\;
\tilde{C}(\omega)}{\tilde{C}_{incoh}(\Omega)},
\end{equation}
where $\tilde{H}(\omega)$ denotes the Fourier cosine transform of
$H(\tau)$, i.e., $\tilde{H}(\omega)=2/\pi \int_0^\infty \ud\tau\,H(\tau)
\cos (\omega \tau)$. Note that this definition of the output SNR differs
by a factor $2$, stemming from the same contribution at $\omega = -
\Omega$, from the definitions used in earlier works \cite
{RMP,PRA91}.
The periodicity of the coherent part gives rise to delta
peaks in the spectrum. Thus, the only contribution to the numerator in
Eq.\ (\ref{snr}) stems from the coherent part of the correlation
function. The output SNR can then be expressed as
\begin{equation}
\label{snr1}
R_{out}=\frac{Q_u}{Q_l},
\end{equation}
where
\begin{equation}
\label{num}
Q_u= {\frac 2T} \int_0^T \ud \tau \,C_{coh}(\tau) \,\cos
(\Omega \tau),
\end{equation}
and
\begin{equation}
\label{den} 
Q_l=\frac 2\pi \int_0^\infty \ud \tau
\,C_{incoh}(\tau) \,\cos (\Omega \tau ).
\end{equation}
A nonmonotonic behavior of the SNR with the strength of the noise is a
signature of the phenomenon of SR.

The signal-to-noise ratio for an input signal $F(t)+\xi(t)$ is given by
\begin{equation}
\label{snrinp}
R_{inp}=\frac{\frac 12 (f_1^2+g_1^2)}{ \frac 2\pi D}.
\end{equation}
where 
\begin{equation}
f_1=\frac{2A}{\pi} \sin \Omega t_c
\end{equation}
and
\begin{equation}
g_1=\frac{2A}{\pi} \left ( 1-\cos \Omega t_c \right )
\end{equation}
It is then clear that for fixed values of $A$ and $D$, $R_{inp}$
increases with $r$ as pointed out in Ref. \cite{ginmak02}. Also,
for given $A$ and $r$, $R_{inp}$ decreases as $D$ increases.
   
The SR gain, $G$, is defined as the ratio of the SNR of the output over
that of the input; namely,
\begin{equation}
\label{gain}
G=\frac {R_{out}}{R_{inp}}.
\end{equation}
SR gain values larger than 1 have been obtained in driven nondynamical
systems \cite{ginmak00}, in stochastic resonators with static
nonlinearities driven by square pulses \cite{chapeau}, or in noisy
bistable systems driven by superthreshold input sinusoidal frequencies
\cite{saga}.  The existence of SR gains with values larger than 1
indicates a truly nonlinear SR.

Although the two-state approximation introduced in Ref. \cite{Casgomprl}
can, in principle, be extended to analyze systems
driven by input signals with $r < 1$, the
analytical expressions obtained are too cumbersome to be of practical
value. Thus, in the present work, we rely on the numerical treatment of
the Langevin equation, Eq.~(\ref{langev}), following the procedure detailed in
Ref.~\cite{casgom03}.

\begin{figure}
\includegraphics[width=9cm,height=10cm]{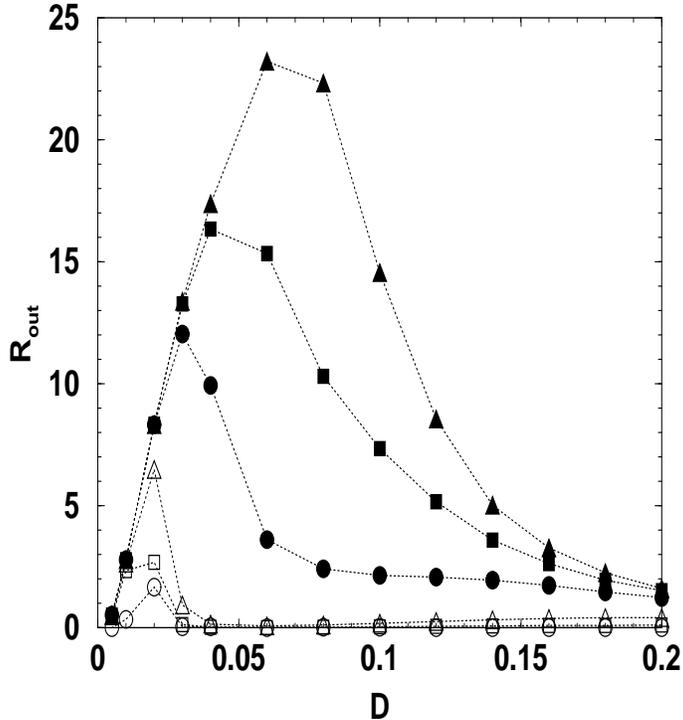}
\caption{\label{SNR} The output signal-to-noise ratio, $R_{out}$ {\it
 vs.} the noise strength, $D$, for $r=0.1$ (circles), $r=0.4$
 (squares), $r=0.7$ (triangles), $r=0.95$ (filled circles),
 $r=0.98$ (filled squares) and $r=1.0$ (filled triangles).
 The input signal has a subthreshold amplitude $A=0.35$ and a
 fundamental frequency $\Omega=2\pi/T=0.0024$.}
\end{figure}

In Fig.\ \ref{SNR}, we depict the behavior of $R_{out}$ with the noise
strength $D$, for input signals of the type given in Eq.~(\ref{pulse}),
with subthreshold amplitude $A=0.35$, fundamental frequency
$\Omega=0.0024$ and $r=0.1, 0.4, 0.7, 0.95, 0.98, 1$. For all values of
$r$, the characteristic nonmonotonic behavior of the SNR with $D$ is
obtained. As $r$ increases, the maximum value of $R_{out}$
increases. Namely, the longer the potential remains asymmetric during
each half-cycle, the larger the maximum height in the SNR is. Therefore,
$R_{out}$ is quite sensitive to the duration of the pulses within each
half-cycle.  

\begin{figure}
\includegraphics[width=9cm,height=10cm]{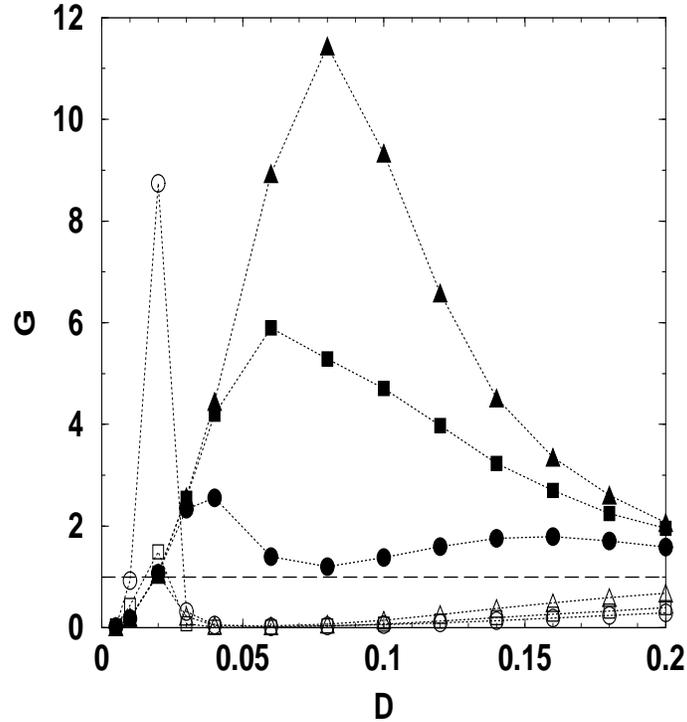}
\caption{\label{Glow} The SR gain, $G$, {\it
 vs.} the noise strength, $D$, for the same parameter values as in Fig.\
 (\ref{SNR}).}
\end{figure}

In Fig.\ \ref{Glow} the behavior of $G$ with the noise strength $D$ is
depicted for the same values of $r$ as in the previous figure. For all
values of $r$ considered, there exists a range of noise values such that
$G$ is larger than unity. The peak value of the SR gain has a
nonmonotonic behavior with $r$.  At the lowest value of $r$ considered,
the SR gain has a rather large peak value. Then, as $r$ is increased,
the peak of the SR gain decreases, in agreement with the observations in
\cite{ginmak01,ginmak02}. As the duration of the pulses gets larger so
that $r$ is closer to $1$, the tendency of the SR gain maximum reverses
and a considerable increase in the maximum is observed.

The just mentioned features can be rationalized by noting that the SR
gain depends on $R_{out}$ and on $R_{in}$. As indicated above, for fixed
values of the noise strength, $D$, and amplitude, $A$, the input
$R_{in}$ always increases with $r$. Also, for given values of $A$ and
$r$, $R_{in}$ decreases monotonically with $D$. On the other hand, the
results depicted in Fig.\ \ref{SNR} indicate that there are two main
effects on the location of the maximum of $R_{out}$ as $r$
increases. First, as noted before, the maximum height increases as $r$
increases. Second, the maxima appear at increasingly large values of $D$
as the duration of the pulses increases. This second effect manifests
itself clearly for pulses of sufficiently long duration, namely for $r$
values larger than $r \approx 0.9$, while it is almost unnoticeable for
smaller bvalues of $r$.  For low values of $r$ (let us say $r =0.1$),
even though the peak of $R_{out}$ is the smallest one appearing in Fig.\
\ref{SNR}, the corresponding value of $R_{in}$ is so small (due to the
smallness of $r$) that the SR gain reaches the large values depicted in
Fig.\ \ref{Glow}. As $r$ increases, the height of the $R_{out}$ maximum
also increases, but appearing at an approximately constant value of the
noise strength. Thus, the increase of $R_{in}$ with $r$ counterbalances
the increase of $R_{out}$ in such a way that the SR gain decreases.
Finally, for long duration pulses, the shift to the right of the
$R_{out}$ maximum and the large increase in its height are the cause of
the increase in the maximum gain observed in Fig.\ \ref{Glow}.

In conclusion, we observe that the behavior of the SR gain for pulses of
relatively short duration is basically a consequence of the behavior of
$R_{in}$, in agreement with the observation of Gingl et
al. \cite{ginmak00,ginmak01,ginmak02}. On the other hand, for $r$ close
to 1, the behavior of the SR gain is dominated by the output SNR.

\begin{acknowledgments}
We acknowledge the support of the Direcci\'on General de
Ense\~nanza Superior of Spain (BFM2002-03822) and the Junta de
Andaluc\'{\i}a.
\end{acknowledgments}

\end{document}